\documentclass[usenatbib,useAMS,letterpaper]{mn2e}
\usepackage{aas_macros}
\usepackage{color}
\usepackage[final]{graphicx}
\usepackage{times}
\usepackage{amssymb}
\usepackage{amsmath}

\newcommand{\Msun}{\ensuremath{\rmn{M}_\odot}}

\title [Maximum stellar mass versus number revisited]
{Maximum stellar mass versus cluster membership number revisited}

\author[Th. Maschberger \& C.J.~Clarke]{Th. Maschberger $^{1,2}$, C.J.~Clarke$^1$\\
$^1$ Institute of Astronomy, Madingley Rd, Cambridge, CB3 0HA, United Kingdom\\
$^2$ Argelander-Institut f{\"u}r Astronomie, Auf dem H{\"u}gel 71, 53225 Bonn, Germany}

\date{Accepted 2008 August 29.  Received 2008 August 29; in original form 2008 July 4}

\pagerange{\pageref{firstpage}--\pageref{lastpage}} \pubyear{2008}

\begin{document}

\label{firstpage}

\maketitle

\begin{abstract}
We have made a new compilation of observations of maximum stellar mass versus cluster membership number from the literature, which we analyse for consistency with the predictions of a simple random drawing hypothesis for stellar mass selection in clusters. Previously,  Weidner and Kroupa have suggested that the maximum stellar mass is lower, in low mass clusters, than would be expected on the basis of random drawing, and have pointed out that this could have important implications for steepening the integrated initial mass function of the Galaxy (the IGIMF) at high masses. Our compilation demonstrates how the observed distribution in the plane of maximum stellar mass versus membership number is affected by the method of target  selection; in particular, rather low $n$  clusters with large maximum stellar masses are abundant in observational datasets that   specifically seek clusters in the environs of high mass stars. Although we do not consider our compilation to be either complete or unbiased, we discuss the method by which such data should be statistically analysed. Our very provisional conclusion is that the data is {\it not} indicating any striking deviation from the expectations of random drawing.
\end{abstract}

\begin{keywords}
stars: mass function - galaxies: stellar content - stars: formation - galaxies: star clusters
\end{keywords}

\section{Introduction}
It is well known (following \citet{weidner+kroupa2004,weidner+kroupa2006} and \citet{oey+clarke2005})  that in the case of clusters containing fewer than $\sim 100$ OB stars (i.e. those with mass $<$ a few $\times~10^4~\Msun$) the maximum stellar mass increases with cluster mass. 
At higher cluster mass scales, the value of the maximum stellar mass saturates at around 150--200~\Msun\  for reasons that are not entirely clear \citep[see e.g.][]{zinnecker+yorke2007}. 
In this paper, we restrict ourselves to considering the lower mass regime. 
In Section 2 we review  why the statistics of maximum stellar masses in clusters of various scales can place constraints on high mass star formation in a cluster context and how rather subtle differences in assumed algorithms for cluster building are imprinted on the integrated galactic IMF (the IGIMF). 
We emphasise that analysis of the statistics of maximum stellar mass versus cluster mass offers the best prospects for an observational determination of whether the IGIMF should be  different from that measured in individual clusters \citep[see e.g.][]{weidner+kroupa2006,elmegreen2006}. 
We also stress that competing algorithms can only be distinguished through proper statistical analysis of the observed distributions and that selecting algorithms according to how they reproduce the {\it mean} trend can be misleading. 
In Section 3 we present a new (but in all likelihood still incomplete and biased) compilation of observational information on this issue and highlight  the sensitivity of the distribution obtained  to the method of target selection. 
In Section 4 we discuss the statistical inferences that can be drawn from the current dataset and conclude (Section 5) with an appeal for further observational information to be used in future analyses.

\section{The importance of maximum stellar mass data and its
statistical analysis}

The simplest interpretation of the fact that the maximum stellar mass is lower, on average, in lower mass clusters is that this just derives from the statistics of random sampling. 
To take a simple analogy, the average height of the tallest inhabitants of large cities is  likely to be greater, on average,  than the average height of the tallest individuals in small villages. 
It would however to be incorrect to infer from this that there is, for example, a nutritional deficiency among village dwellers.
On the other hand, a better analogy might be with the wealth of richest individuals in settlements of various sizes, since in this case this might reflect the size of the local economic base. 
This is the sort of argument used by Weidner \& Kroupa, who point out that in the case of cluster formation, the stars acquire their mass directly from the available gas reservoir. 
Their simple Monte-Carlo simulations build  the finite size of the gas reservoir into their algorithms for stellar mass selection and reject any star whose formation causes the total designated cluster mass to be exceeded. 
This  `rejection' element preferentially affects more massive stars and is chiefly manifest through a statistical lowering of the maximum stellar mass compared with its value in random sampling experiments. 
Another plausible algorithm was proposed by \citet{elmegreen2006}, motivated by the fact that the fraction of the initial gas in a protocluster that ends up in stars may be  significantly less than unity. 
In this algorithm, therefore,  although stellar mass selection is terminated once the total stellar mass exceeds the designated total mass, the last star is only rejected if this takes the total cluster mass over a value equal to the sum of the designated cluster mass and the mass of an additional gas reservoir. 
Since this is a softer rejection criterion, this algorithm produces results that are closer to random sampling than a strictly mass constrained algorithm.

\begin{figure}
\includegraphics[width=8.5cm]{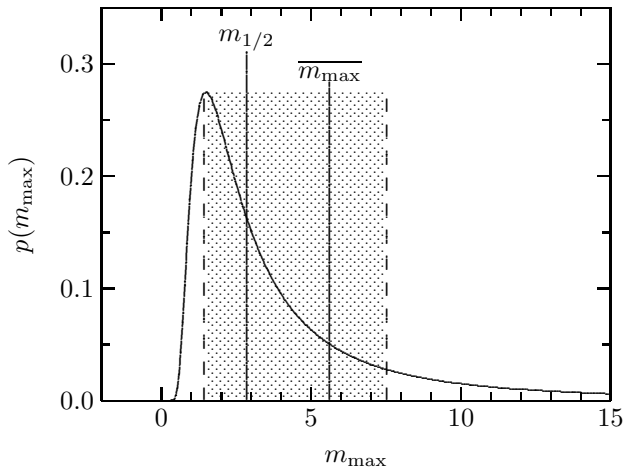}
\caption{\label{mmaxpdfplot}
Probability density of the most massive star, $p(m_\mathrm{max})$ (eq. \ref{mmaxpdf} in Appendix \ref{distmmax}) for a star cluster containing $n=30$ stars.
Characteristic quantities are the mean, $\overline{m_\mathrm{max}}$, and the median, $m_{1/2}$.
The $1/6$th and $5/6$th quantiles limit the shaded region containing $2/3$rd of the most massive stars.
}
\end{figure}

Since in all these  cases the maximum stellar mass follows quite a broad distribution (at fixed cluster scale), these differences cannot be discerned from a single observational datapoint. 
Instead it is necessary to compare observed distributions (at given cluster scale) with the results of simulations (or, in the case of the simplest, random drawing, hypothesis, with the results of analytic predictions). 
Some care is needed when considering the best property of the distribution that should be compared with observational data, as is demonstrated in Figure \ref{mmaxpdfplot} which illustrates the probability density function of maximum stellar mass at given cluster mass scale in the case of the random drawing hypothesis. 
Clearly, this distribution is highly asymmetric, with the mean significantly exceeding the median. 
This means that in the case of sparsely sampled observational data (i.e. not many clusters at given mass scale), the observed mean is likely to be lower than the true mean. 
A better approach is instead to compute the predicted distribution (as a function of cluster mass) and use a non-parametric method (e.g. a Kolomogorov-Smirnov test) to compare these with the  distribution of observational datapoints in the plane of cluster mass versus maximum stellar mass (see \citet{oey+clarke2005} and Section 3).

Thus far we have discussed the interpretation of these statistics in terms of what  light they may shed on cluster formation (obviously the algorithms described above are simple `toy models' but a clear signature in favour of one of them could be useful, for example, in determining how much stellar mass assignments are shaped by strict limitations in available gas supply). 
Another implication is purely empirical: we have stressed that the steepening of the upper IMF (and the resulting reduction in maximum stellar mass) in the case of non-random mass selection algorithms are too subtle to be detectable in any given cluster (i.e. each cluster is statistically consistent with being drawn from the input  IMF). 
However, when one combines the results of many clusters (i.e. - on the assumption that  the galactic field is composed of dissolved clusters  - if one turns an IMF into an IGIMF) the  signature of algorithms that preferentially reject high mass stars is seen in a steepening of the IGIMF. 
This important insight was first discussed in this way by \citet{kroupa+weidner2003} (though see \citet{vanbeveren1982} for an earlier version of the argument). 
The reason why such star rejection algorithms - which are only important in the lower mass clusters that we discuss here -  have a discernible effect on the IGIMF is simply that, given the steepness of the observed cluster mass function,  low mass clusters make an important contribution to the galactic field.

Naturally, a steeper IGIMF has implications for  how a range of quantities (such as supernova rate or ionising luminosity or chemical enrichment) relate to the galactic star formation rate. 
\citet{weidner-etal2004} also extended the argument by positing that similar considerations  apply to {\it star cluster} maximum masses in galaxies of different masses. 
This means that in lower mass galaxies, the field population would be more dominated by lower mass clusters, and therefore that the IGIMF would be more steepened by the effect described above. 
Indeed, \citet{pflammaltenburg-etal2007} have gone on to argue that this has important implications for the mapping between H$\alpha$ luminosity and star formation rate in dwarf galaxies and would lead to the systematic {\it under-}estimation of the SFR in dwarfs. 

Observationally, opinion is strongly divided as to whether there is good evidence that the IGIMF is  steeper than Salpeter (or if it varies between galaxies): see e.g. \citet{elmegreen2006}, \citet{pflammaltenburg-etal2007}, \citet{selman+melnick2008astroph} and discussion in \citet{clarke2008}.
We may, however, be able to turn this question around: {\it if} we can use cluster data to determine whether the maximum stellar mass statistics are indeed compatible with random drawing models then we can immediately learn  whether the IGIMF should be equal to the IMF (without recourse to any Galaxy-wide or extragalactic data). 
Although, as we shall see in the following Section, it is not straightforward to achieve an unbiased sample for analysis, it is obviously attractive to be able to use rather simple, local observations to constrain a quantity which is potentially of extragalactic significance.

\section{Observational data}

\begin{figure*}
\includegraphics[width=16cm]{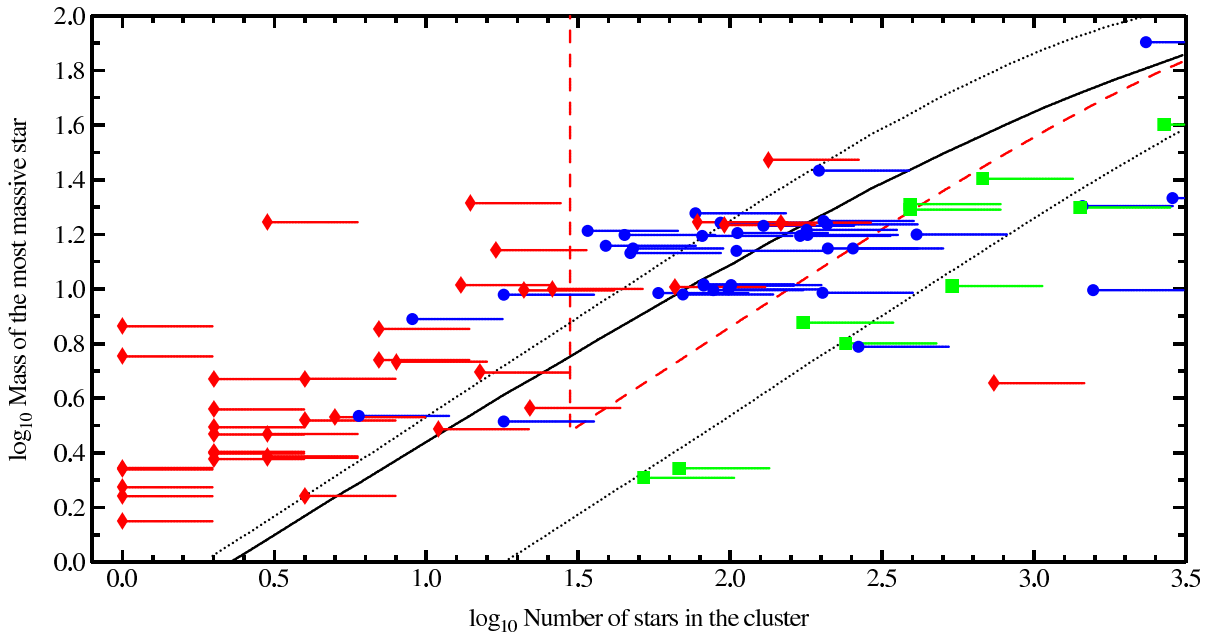}
\caption{\label{mmaxntot}
Mass of the most massive star versus the number of stars in the cluster (for  better visibility, a small random scatter was applied to the (discrete) masses).
The data are collected from the literature, with the main sources Testi et al. ({\color{red} $\blacklozenge$}) and Weidner \& Kroupa ({\color{green} $\blacksquare$}).
The references for the other points are given in Appendix \ref{datatable}.
The solid line is the mean value of $m_\mathrm{max}$ depending on $n$.
The dotted lines follow the $1/6$ and $5/6$ quantiles, and should confine $2/3$rd of the observed data.
}
\end{figure*}

In what follows, we attempt a simple test and enquire: is observational data on maximum stellar mass as a function of cluster scale compatible with the hypothesis of random drawing from a universal IMF (i.e. the same exponent and mass limits for all clusters)?  Incompatibility would have important consequences for the IGIMF, as we have seen above and we would then need to enquire what other algorithms could achieve consistency. 
On the other hand, compatibility (at whatever desired significance level) would not necessarily  imply that random drawing is the `best fit model'; it would however remove much of the motivation for finding more complex alternatives. 

The analytical model for random drawing of stellar masses (see  \citet{oey+clarke2005}  or \citet{selman+melnick2008astroph},  and Appendix \ref{distmmax} of this paper that recapitulates the main results) is based on the expected distribution of maximum stellar masses in the case that one makes a given number, $n$, of selections from a given mass function. 
It therefore makes sense to make $n$ the independent variable (unlike in the case of mass constrained models where cluster mass is the obvious choice). 
In order to obtain a homogeneous sample the data are renormalised to a common lower limit because of differing observational lower mass limits for each cluster.
Each cluster is designated by the expected number of stars that it would contain down to $0.08~\Msun$ (i.e. not including brown dwarfs), given the observed number and mass limit, and assuming an IMF for the missing range.\footnote{Note that the choice of this lower limit is arbitrary, provided that it is self-consistently applied  to all the observational data and to the analytic predictions.}
We employ a two-part power law IMF \citep{kroupa2001,kroupa2002} with a scaling as $m^{-2.35}$ \citep{salpeter1955} for stars $>~1~\Msun$.
In order to compare the completeness magnitude of a particular set of observations with the lower mass limit of our analysis, we use the conversion between K magnitude and mass given by \citet{carpenter-etal1993} ($ m = 10^{-0.24 m_K + 0.24} $ or $ m = 10^{-0.25 m_H + 0.44} $).
All clusters are corrected (at least in a statistical sense) for background or foreground contamination.
Furthermore we demanded that the observed region was large enough to contain the whole cluster area.

As far as the maximum stellar mass is concerned, we either use values quoted in the literature or else estimate masses from listed spectral types using the conversion given in \citet{schmidt-kaler1982} (the masses for spectral types not contained in the list being interpolated).

Our criterion for including a cluster is only that we have found it to be possible to derive estimates of both $n$ and $m_\mathrm{max}$ in this way. 
As we discuss below, it is unlikely to be either a complete or an unbiased sample and this makes any conclusions that we draw from this dataset extremely preliminary. 
Figure \ref{mmaxntot} compares all the data that we have assembled with the predicted centiles of the random drawing model (i.e.the mean and the $1/6$ and $5/6$ contours of the cumulative distribution). 
The data is coded according to source: {\color{green}$\blacksquare$} for the data tabulated in  Weidner \& Kroupa,{\color{red}$\blacklozenge$} for that obtained by Testi et al and {\color{blue}\LARGE\textbullet} for miscellaneous other observations (see Appendix \ref{datatable}). 

One of the hardest aspects of constructing Figure \ref{mmaxntot} is the assignment of realistic errorbars (in $n$; errors in $m_\mathrm{max}$ are negligible by comparison, since we include only clusters which are young enough for their most massive members not to have expired as supernovae). 
We have drawn one-sided errorbars, on the grounds that  we are probably missing stars that are located at large distances from the most massive star where the density of sources on the sky falls below the local background value, either as a result of initial conditions or dynamical evolution.
We are interested here in the total population of stars that was formed with the most massive object, irrespective of whether these stars are currently bound to the natal cluster. 
Dynamical evolution in small $n$  clusters can however cause significant expansion over a few Myr \citep{bonnell+clarke1999}  and  this effect increases the likelihood that we may be missing stars at large distances. 
In order to estimate the possible error introduced in this way, we really need dynamical simulations on a cluster by cluster basis, which limit the range of original configurations (cluster $n$ and size) that are compatible with the present census of background corrected objects. 
To our knowledge, this exercise has only been undertaken in one cluster ($\eta$ Cha, \citealp{moraux-etal2007}) where the total  $n$ lies in the range of $18$ (as observed) to $40$ (the maximum number that is compatible with leaving a cluster with the parameters observed). 
With this in mind, we add one-sided errorbars of a factor of $2$ in Figure \ref{mmaxntot}, although note that this is pessimistic (i.e. too large) for the larger $n$ clusters where the  rate of dynamical dispersion is probably lower.

An obvious feature of Figure \ref{mmaxntot} is that different regions of the $n,m_\mathrm{max}$ plane are populated by clusters obtained through different observing strategies. 
Evidently, the squares are low compared with the centiles, explaining why Weidner \& Kroupa found it necessary to invoke a non-random algorithm for stellar mass selection. 
The data of Testi et al. is apparently discrepant in the opposite direction - i.e. maximum stellar masses are, if anything, rather large, given the number of stars in the clusters. 
The reason why these two datasets are complementary (and both necessary to the statistical analysis) is simply one of the order in which the properties of the systems were determined. 
Weidner \& Kroupa sought data on  regions recognised as `star clusters' and then found the maximum recorded mass. 
Testi et al. instead first identified massive stars (including those that are apparently isolated) and then undertook deep infrared imaging of the environs in order to identify any surrounding over-density of low mass stars. 
Unsurprisingly, the ratio of maximum stellar mass to cluster number is considerably higher in the latter case.      

\section{Analysis of the dataset}\label{analysisdataset}

The observational data contained in Figure \ref{mmaxntot} is highly incomplete  and biased and so great care must be taken in its statistical analysis. 
In this section we discuss whether  a subset of the data can be used to settle whether the results  are consistent with the expectations of the random drawing hypothesis. 
We here remind the reader that acceptance (rejection) of this hypothesis means that the high mass tail of the IGIMF should be identical to (steeper than) the input Salpeter IMF.

If we simply took all these datapoints at face value, we could  evaluate a cumulative probability  for each datapoint (i.e. evaluate the probability that the maximum mass is less than or equal to the datapoint value, according to  the theoretical distribution for that particular value of $n$): 
if the observations match theoretical expectations, these probabilies should be uniformly distributed between $0$ and $1$. 
We can then use a KS test to compare the probability distribution with a  uniform distribution. 
The probability that data generated through random drawing would be as discrepant from the theoretical prediction as is that observed is $10^{-17}$ (adopting the membership numbers denoted by symbols in Figure \ref{mmaxntot}) and $10^{-8}$ (if one instead adopts twice these values, i.e. corresponding to the upper end of the errorbars shown in Figure \ref{mmaxntot}). 
At face value, therefore, one would overwhelmingly reject the hypothesis of random drawing. 
The reason for the discrepancy (as can be seen from  Figure \ref{mmaxntot}) is that the lower range of the cumulative distribution function is actually {\it under-}populated by the data.

This conclusion is however highly misleading - the discrepancy is strongly driven by the very large number of  datapoints in Figure \ref{mmaxntot} (from Testi et al.) which populate the upper regions of the cumulative distribution. 
However, it needs to be remembered that we simply do not have complete data.

One way forward is to define a stellar mass, $m_\mathrm{cpl}$, such that we deem that we have the information on all clusters (of all $n$), for which the maximum mass is $> m_\mathrm{cpl}$. 
By retaining all the data with $m_\mathrm{max} > m_\mathrm{cpl}$, however, we then have data which is complete down to different  values  of the cumulative distribution depending on the value of $n$, which is impractical for a statistical test and its interpretation. 
A more convenient approach is to select the data such that they are complete down to the same cumulative probability $P_\mathrm{cut}$.
By this criterion all data points are selected which lie above the mass $m_\mathrm{cut} (n)$ which corresponds to $P_\mathrm{cut}$ in the cumulative distribution of $m_\mathrm{max} (n)$.

Furthermore, we do not take into account the very small $n$ data, because of their presumably large error bars.
Thus we introduce a minimum number of stars in a cluster, $n_\mathrm{cpl}$, which are needed to ensure the quality of the selected data.
The selection parameter $P_\mathrm{cut}$ follows then as the cumulative probability of $m_\mathrm{cpl}$ for the given $n_\mathrm{cpl}$.
The choice for the values of $m_\mathrm{cpl}$ and $n_\mathrm{cpl}$ is discussed below.

We demonstrate the selection criterion in Figure \ref{mmaxntot} for the case $m_\mathrm{cpl} = 3 \Msun, n_\mathrm{cpl} = 30$, where the diagonal dashed line tracks the values of  $m_\mathrm{cut} (n)$ with  $P_\mathrm{cut} =0.53$.
By restricting ourselves to data within the wedge of the diagram above the dashed line, we are obviously not using a lot of the data, but have now defined a sample which is 
complete down to a fixed point in the cumulative distribution  at every value of $n$ included. 
We can therefore test whether these cumulative probabilities  are uniformly  distributed in the range $P_\mathrm{cut}$ to $1$. 

If we, for the moment, disregard any physical or observational ground for choosing particular values of $m_\mathrm{cpl}$ and $n_\mathrm{cpl}$, we can use different values of these parameters so as to explore various aspects of the two dimensional distribution. 
As expected, the KS probabilities are sensitive to the values of the $m_\mathrm{cpl}, n_\mathrm{cpl}$ adopted, with probabilities being higher if the choice is such (e.g. through high $m_\mathrm{cpl}$ or either very high or very low $n_\mathrm{cpl}$) that the number of datapoints retained is small. 

\begin{figure}
\includegraphics[width=8.5cm]{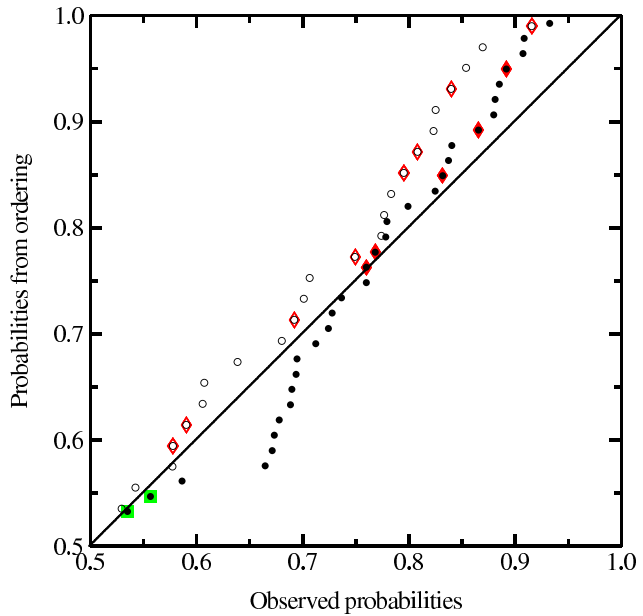}
\caption{\label{quantiles}
A plot of the ordered cumulative probabilities versus the observed cumulative probabilities 
derived from the distribution of $m_\mathrm{max}$ for a particular $n$.
The shown data fullfill the selection criterion discussed in Section \ref{analysisdataset}, i.e. they  lie in the dashed wedge in Fig. \ref{mmaxntot}.
The filled symbols follow  with the observed $n$, whereas for the open symbols $2n$ was used, corresponding to the right end of the error bars in Fig. \ref{mmaxntot}  ({\color{green}$\blacksquare$}: Weidner \& Kroupa; {\color{red}$\blacklozenge$} Testi et al.).
If the data were uniformly distributed they should follow the diagonal.
}
\end{figure}

We are particularly looking for evidence for a lack of data at the upper end of the cumulative distribution (as this would imply $m_\mathrm{max}$ was less than that implied by random drawing); 
we are however finding in general that this is not a striking feature of the distribution. 
In fact, a significant dearth of high $m_\mathrm{max}$ values is only recorded if one selects $m_\mathrm{cpl}$ and $n_\mathrm{cpl}$ so as to just include the clump of datapoints  at around $10 \Msun$ and $n \sim 60$, since in this case the very top of the cumulative distribution is slightly under-represented  (for example, for $m_\mathrm{cpl} = 9 \Msun$ and $n_\mathrm{cpl} = 60$, there are many more datapoints with cumulative probabilities  slighly larger than $P_\mathrm{cut}$ ($=0.75$) compared with those at higher values, and this is reflected in a relatively low KS probability of around $2 \%$). 
In general, however, the feature of the plot that is generally flagged as most discrepant by the KS test is {\it not} a dearth of the highest values, but instead the lack of systems at around $3 \Msun < m_\mathrm{max} < 10 \Msun$ and $n \sim  50-100$, which is  readily visible as a data hole in Figure \ref{mmaxntot}. 
 
It is obviously unsatisfactory if we tune the values of $m_\mathrm{cpl}$ and  $n_\mathrm{cpl}$ so as to retroactively highlight a particular aspect of the data. 
Therefore, as our best guess of plausible parameter values,  we adopt $m_\mathrm{cpl} = 3 \Msun$ and $n_\mathrm{cpl} = 30$. 
The choice of $n_\mathrm{cpl}$ is motivated by the fact that we consider the errorbars in membership number of smaller clusters to be very high: both because dynamical evolution is more rapid in smaller $n$ systems \citep{bonnell+clarke1999} and also because the ejection of even a few stars to radii where they cannot be distinguished from the background causes relatively large fractional errors in $n$. 
The choice of $m_\mathrm{cpl}$ reflects the mass of a moderately luminous Herbig Ae star as targeted by Testi et al 1998. 
We have cautioned above that we must not go to much lower masses, since young stars of close to solar mass and below have not been systematically targeted for surrounding clusters.

 The distributions derived from the data, for this choice of $m_\mathrm{cpl}$ and
$n_\mathrm{cpl}$, are represented graphically in  Figure \ref{quantiles}. 
For each retained datapoint,  we calculate the position in the theoretical cumulative distribution function (plotted on x-axis) and on the y-axis we plot the ranked position of the datapoint. 
(Note that this latter quantity has been renormalised to lie in the interval $[P_\mathrm{cut},1]$ instead of $[0,1]$)
The filled symbols use membership numbers denoted by the symbols in Figure 2, whereas the  open symbols correspond to the case when values of $n$ a factor two larger are adopted. 
Note that the number of datapoints are not the same in the two cases, since shifting $n$ by a factor of two moves data values in and out of the region above the dashed  line in Figure 2.
If the data conforms to the random hypothesis, then the data plotted in Figure 3 should be following the diagonal.

A KS test performed on this data yields a KS probability of nearly $20 \%$, implying quite adequate agreement with the random drawing hypothesis
\footnote{We also tested the sensitivity of our results to the mapping employed between spectral type and mass by noting that if one uses the calibration of \citet{martins-etal2005}, two of the O-stars in our sample are significantly reduced in mass. 
This adjustment however does not change the reasonable agreement with the null hypothesis.}.
Although both curves in Figure 3 are somewhat above the diagonal at the uppermost end  (implying a mild deficit of data values at the top of the predicted cumulative distribution function), this discrepancy  is not significant in this sample (and, as discussed above,  the feature in the filled curve that is  most discrepant is actually the mild deficit of    data around $\sim 0.6$ in the cumulative distribution, corresponding to the `data hole' at $m_\mathrm{max} \sim 3-10 \Msun, n \sim 50-100$ in Figure 2).
  
\section{Conclusions}

We have high-lighted the difficulty in analysing the data contained in Figure \ref{mmaxntot} owing to difficulties to assigning regions of the diagram where the data is believed to be complete. 

Nevertheless, our preliminary conclusion is that we are {\it not} seeing strong evidence for a systematic suppression in maximum stellar mass in small $n$ clusters in addition to that expected on the basis of the statistics of random drawing (see also the complementary analysis of the statistics of isolated stars by \citet{parker+goodwin2007}, which reached similar conclusions). 
Indeed, if anything, the feature of Figure \ref{mmaxntot} that seems to be most discrepant with the random drawing model is the data hole in the range $m_\mathrm{max} \sim 3-10 \Msun, n \sim 50-100$. 
We are however aware that this might indeed be filled in if we have under-estimated the incompleteness in smaller $n$ clusters (particularly  due to the effects of dynamical evolution).

Our conclusion (in support of the random drawing hypothesis) remains provisional. 
Although we have set out what we believe to be a statistically correct methodology for analysing the problem, we are highly aware of the difficulties of properly quantifying observational selection effects. 
We therefore seek further input from observers in compiling a good sample for this kind of analysis.

\section{Acknowledgements}
We thank Carsten Weidner, Leonardo Testi, Thomas Preibisch, Pavel Kroupa and the referee Hans Zinnecker for useful discussions and helping to collect the data.
ThM acknowledges financial support via an EARA-EST fellowship and the European Union Research Training Network ``Constellation''.

\bibliographystyle{mn2e}
\bibliography{cluster,clusterdata}

\appendix

\section{The distribution of $m_\mathrm{max}$}\label{distmmax}
The IMF used in this work is according to \citet{kroupa2001,kroupa2002} with a Salpeter exponent for massive stars \citep{salpeter1955,massey1998},
\begin{eqnarray} \xi (m) &\propto& \begin{cases} 
m^{-1.3} & m_\mathrm{MIN}  \le m < 0.5\ \Msun \\
m^{-2.35} & 0.5\ \Msun \le m < m_\mathrm{MAX} \\
\end{cases}, \end{eqnarray}
where the lower limit $m_\mathrm{MIN} = 0.08\ \Msun$ and brown dwarfs are not included.
Since we consider `pure' random sampling the upper limit does not depend on the total mass of the star cluster ($m_\mathrm{MAX} = 150\ \Msun$).
We use the IMF as a probability density, i. e. normalised to
\begin{eqnarray} \int_{m_\mathrm{MIN}}^{m_\mathrm{MAX}} \xi (m) \mathrm{d} m &=& 1. \end{eqnarray}
In a sample of ``identical'' star clusters (with the same $n$) the mass of the most massive star in each cluster will not be the same but follow its own distribution function.
For this kind of Monte-Carlo experiment the distribution of the most massive star can analytically be derived.
The probability for the most massive star to lie in the mass interval $m_\mathrm{max},m_\mathrm{max}+\mathrm{d} m$ is 
\begin{eqnarray} P ( m \in [m_\mathrm{max},m_\mathrm{max}+\mathrm{d} m]) = \xi (m_\mathrm{max}) \mathrm{d} m.  \label{pmmax} \end{eqnarray}
All other stars must have a mass smaller than $m$.
The probability to pick randomly $n-1$ stars from the mass range $m_\mathrm{MIN},m$ is
\begin{eqnarray} P (m_{1\dots n-1} \in [m_\mathrm{MIN},m_\mathrm{max}]) &=& \left( \int_{m_\mathrm{MIN}}^{m_\mathrm{max}} \xi (m') \mathrm{d} m') \right)^{n-1} \label{prest} \end{eqnarray}
The probability distribution  of the most massive star is then the product of eqns. \ref{pmmax} and \ref{prest}, multiplied with the factor $n$ because every star could be the most massive star.
To obtain the probability distribution the product has to be differentiated with respect to $m$.
This gives
\begin{eqnarray} p (m_\mathrm{max}) &=& n \left( \int_{m_\mathrm{MIN}}^{m_\mathrm{max}} \xi (m') \mathrm{d} m' \right)^{n-1} \xi (m_\mathrm{max}) \label{mmaxpdf} \end{eqnarray}

\section{Data of the used clusters}\label{datatable}
In the compilation of \citet{weidner+kroupa2006} the total mass of a cluster down to $0.01~\Msun$ is given.
We converted this mass into a membershp number by dividing with the average stellar mass ($0.36~\Msun$), and subtracted the expected number of brown dwarfs ($\xi (m) \propto m^{-0.3}$ for $0.01\ \Msun  \le m < 0.08\ \Msun$)
.
In the cases where numbers are given we used them.

From the list of Testi et al. we took the $I_C$ values, which are background-corrected total numbers and range down to the limiting magnitude $\mathcal{M}_\mathcal{K}^c$.
One particular star/cluster, MWC 300 ($m_\mathrm{max} \approx 5~\Msun$), is quite distant ($15.5$~kpc) and has a very high limiting mass ($\approx~2.4~\Msun$). 
Therefore the observed number of $21$ stars is corrected to $740$ stars.
This is the far right outlier in Fig. \ref{mmaxntot}, and the correctness of this data point is questionable, but included to be complete.

\begin{tabbing}
 IRAS 123456789012345 	\= O1234 			   	  	\= 123456789 			\= 123 \= \kill 
 Name  			\>\> $m_\mathrm{max}$ \' \> $ n $ \' \> Ref 	\\  
 Taurus-Auriga \> \>       2.2 \' \> 68 \' \> 1 \\ 
Ser SVS 2 \> \>       2.2 \' \> 52 \' \> 1 \\ 
$\rho$ Ophiuchi \> \>       8.0 \' \> 174 \' \> 1 \\ 
IC 348 \> \>       6.0 \' \> 241 \' \> 1 \\ 
IC 348 \> \>       5.9 \' \> 265 \' \> 2 \\ 
 NGC 2024  \> \>      20.0 \' \> 1447 \' \> 3 4 \\ 
 NGC 2024  \> \>      20.0 \' \> 392 \' \> 1 \\ 
$\sigma$ Orionis \> \>      20.0 \' \> 392 \' \> 1 \\ 
Mon R2 \> \>      10.0 \' \> 538 \' \> 1 \\ 
Mon R2 \> \>      10.0 \' \> 1568 \' \> 5 6 \\ 
NGC 2264 \> \>      25.0 \' \> 679 \' \> 1 \\ 
 NGC 6530  \> \>      20.0 \' \> 1421 \' \> 1 \\ 
 NGC 6530  \> \>      80.0 \' \> 2337 \' \> 7 \\ 
Ber 86 \> \>      40.0 \' \> 2682 \' \> 1 \\ 
USco \> \>      22.0 \' \> 2859 \' \> 8 \\ 
 IRAS 00494+5617  \> \>      14.0 \' \> 105 \' \> 9 10 \\ 
 IRAS 02575+6017  \> \>      14.0 \' \> 210 \' \> 11 \\ 
 IRAS 02575+6017  \> \>      14.0 \' \> 254 \' \> 9 10 \\ 
 IRAS 02593+6016   \> \>      19.0 \' \> 77 \' \> 11 \\ 
 IRAS 02593+6016   \> \>      17.5 \' \> 93 \' \> 9 10 \\ 
IRAS 03064+5638 \> \>      16.0 \' \> 34 \' \> 9 10 \\ 
 IRAS 05100+3723  \> \>      17.5 \' \> 203 \' \> 9 10 \\ 
 IRAS 05197+3355  \> \>      16.0 \' \> 170 \' \> 9 10 \\ 
 IRAS 05274+3345  \> \>      10.0 \' \> 18 \' \> 9 10 \\ 
 IRAS 05275+3540  \> \>      16.0 \' \> 179 \' \> 9 10 \\ 
 IRAS 05377+3548   \> \>      14.0 \' \> 39 \' \> 9 10 \\ 
 IRAS 05490+2658   \> \>      10.0 \' \> 58 \' \> 9 10 \\ 
IRAS 05553+1631 \> \>      10.0 \' \> 82 \' \> 9 10 \\ 
 IRAS 06056+2131  \> \>      10.0 \' \> 202 \' \> 9 10 \\ 
 IRAS 06058+2138  \> \>      10.0 \' \> 99 \' \> 9 10 \\ 
 IRAS 06068+2030  \> \>      16.0 \' \> 81 \' \> 9 10 \\ 
 IRAS 06073+1249  \> \>      16.0 \' \> 412 \' \> 9 10 \\ 
 IRAS 06155+2319  \> \>      14.0 \' \> 48 \' \> 9 10 \\ 
 IRAS 06308+0402   \> \>      16.0 \' \> 45 \' \> 9 10 \\ 
MWC 1080 \> \>      17.5 \' \> 78 \' \> 12 \\ 
 $\eta$ Cha  \> \>       3.5 \' \> 18 \' \> 13 \\ 
 M20  \> \>      27.0 \' \> 196 \' \> 14 \\ 
IRAS 01546+6319 \> \>      14.0 \' \> 47 \' \> 11 \\ 
IRAS 02044+6031 \> \>      17.5 \' \> 129 \' \> 11 \\ 
IRAS 02232+6138 \> \>      16.0 \' \> 180 \' \> 11 \\ 
IRAS 02245+6115 \> \>      16.0 \' \> 106 \' \> 11 \\ 
IRAS 02461+6147 \> \>      10.0 \' \> 101 \' \> 11 \\ 
MWC 137 \> \>      17.5 \' \> 96 \' \> 12 15 \\ 
MWC 297 \> \>      21.0 \' \> 14 \' \> 12 \\ 
 NGC 7129  \> \>      10.0 \' \> 66 \' \> 12 \\ 
 NGC 7129  \> \>      10.0 \' \> 70 \' \> 16 \\ 
 NGC 7129  \> \>      10.0 \' \> 82 \' \> 17 \\ 
 NGC 7129  \> \>      10.0 \' \> 88 \' \> 18 \\ 
MaC H12 \> \>       2.0 \' \> 4 \' \> 12 \\ 
VX Cas \> \>       2.9 \' \> 4 \' \> 12 \\ 
RNO 1B \> \>       5.0 \' \> 8 \' \> 12 \\ 
XY Per \> \>       5.2 \' \> 7 \' \> 12 15 \\ 
MWC 480 \> \>       2.5 \' \> 3 \' \> 12 \\ 
HD 245185 \> \>       2.7 \' \> 3 \' \> 12 \\ 
HD 37490 \> \>       7.6 \' \> 7 \' \> 12 15 \\ 
VY Mon \> \>       3.8 \' \> 22 \' \> 12 \\ 
VV Ser \> \>       3.5 \' \> 11 \' \> 12 \\ 
LkH$\alpha$ 257 \> \>       3.8 \' \> 5 \' \> 12 \\ 
HD 216629 \> \>      10.0 \' \> 26 \' \> 12 \\ 
BD+40 4124 \> \>      10.0 \' \> 13 \' \> 12 15 \\ 
V645 Cyg \> \>      30.0 \' \> 134 \' \> 12 \\ 
IP Per \> \>       2.3 \' \> 3 \' \> 12 \\ 
MWC 300 \> \>       5.0 \' \> 739 \' \> 12 \\ 
AS 310 \> \>      17.5 \' \> 147 \' \> 12 \\ 
HD 200775 \> \>       7.6 \' \> 1 \' \> 12 \\ 
LkH$\alpha$ 233 \> \>       1.6 \' \> 1 \' \> 12 \\ 
Elias 1 \> \>       1.9 \' \> 1 \' \> 12 \\ 
V1012 Ori \> \>       3.5 \' \> 2 \' \> 12 \\ 
MWC 758 \> \>       2.3 \' \> 2 \' \> 12 \\ 
RR Tau \> \>       2.3 \' \> 1 \' \> 12 \\ 
LkH$\alpha$ 208 \> \>       2.3 \' \> 2 \' \> 12 \\ 
BHJ 71 \> \>      17.5 \' \> 3 \' \> 12 \\ 
LkH$\alpha$ 215 \> \>       4.6 \' \> 4 \' \> 12 \\ 
HD 97048 \> \>       3.4 \' \> 6 \' \> 17 \\ 
BD+46 3474 \> \>      17.5 \' \> 209 \' \> 17 \\ 
BD+46 3471 \> \>       7.3 \' \> 9 \' \> 17 \\ 
RNO 6 \> \>      14.0 \' \> 17 \' \> 12 15 \\ 
HD 52721 \> \>      10.0 \' \> 21 \' \> 12 15 \\ 
HD 259431 \> \>       5.9 \' \> 1 \' \> 12 15 \\ 
LkH$\alpha$ 25 \> \>       4.6 \' \> 15 \' \> 12 15 \\ 
HD 250550 \> \>       4.6 \' \> 2 \' \> 12 15 \\ 
LkH$\alpha$ 218 \> \>       3.5 \' \> 2 \' \> 12 15 \\ 
AB Aur \> \>       2.9 \' \> 2 \' \> 12 15 \\ 
T Ori \> \>       2.3 \' \> 1 \' \> 12 15 \\ 
HK Ori \> \>       2.1 \' \> 2 \' \> 12 15 \\ 
BF Ori \> \>       1.6 \' \> 1 \' \> 12 15 \\ 
\end{tabbing} References: 1: \citet{weidner+kroupa2006};
2: \citet{luhman-etal2003};
3: \citet{lada-etal1991};
4: \citet{bik-etal2003a};
5: \citet{carpenter-etal1997};
6: \citet{carpenter2000};
7: \citet{prisinzano-etal2005};
8: \citet{preibisch-etal2002};
9: \citet{carpenter-etal1993};
10: \citet{carpenter-etal1990};
11: \citet{carpenter-etal2000};
12: \citet{testi-etal1998};
13: \citet{moraux-etal2007};
14: \citet{rho-etal2001};
15: \citet{testi-etal1997};
16: \citet{gutermuth-etal2004};
17: \citet{wang+looney2007};
18: \citet{gutermuth-etal2005};

\label{lastpage}
\end{document}